\documentclass[prb]{revtex4}
\usepackage{graphicx}% Include figure files
\usepackage{dcolumn}% Align table columns on decimal point
\usepackage{bm}% bold math

\begin{document}

\title{Mott-Hubbard localization in model of electronic subsystem of doped fullerides}

\author{Yu.~Dovhopyaty, L.~Didukh, O.~Kramar, Yu.~Skorenkyy, Yu.~Drohobitskyy}
\affiliation{Ternopil National Technical University, 56, Ruska Str., Ternopil, 46001, Ukraine}

\date{\today}% It is always \today, today,
             %  but any date may be explicitly specified

\begin{abstract}
Microscopical model of a doped fulleride electronic subsystem taking into account the triple orbital degeneracy of energy states is considered within the configurational-operator approach. Using the Green function method the energy spectrum of the model at integer band filling $n=1$ is calculated, which case corresponds to $AC_{60}$ compounds. Possible correlation-driven metal-insulator transition in the model is discussed.
\end{abstract}

\maketitle

\section{Introduction}

Electrical, optical and mechanical properties of fullerenes~\cite{elec95,mani06} in condensed state demonstrate
considerable physical content of phenomena which take place in fullerenes
and show that the use of such materials in electronics has significant perspectives. Fullerene crystals and films are semiconductors with an energy gap of $1.2-1.9 eV$~\cite{sait91,achi91} and have photoconductivity under visible light irradiation. Fullerene crystals have comparatively small binding energy and at room temperature the phase transition
connected with orientational disordering of fullerene molecules take place in such crystals~\cite{heyn91}.

Addition of radicals containing platinum group metals~\cite{hawk92} to fullerenes C$_{60}$ allows
to obtain ferromagnetic material based on fullerene. In polycrystal C$_{60}$ doped by alkali metal superconductivity at temperature lower then $33 K$ is observed~\cite{flem91,holc91}. Large binding energy is typical for metallocarbohedrenes M$_{8}$C$_{12}$, where $M=Ti, V, Hg, Zr$. For example, in Ti$_{8}$C$_{12}$ molecule binding energy per atom is $6.1 eV$ ~\cite{redd92} (for C$_{60}$ molecule this energy is $7.4-7.6 eV$~\cite{sait91}).

Fullerenes in solid state (fullerites) are the molecular crystals, where interaction between atoms in C$_{60}$ molecule is much larger then interaction between nearest molecules. In tightly packed structure each fullerene molecule has $12$ nearest neighbors. Depending on peculiarities of molecular interaction, face-centered cubic lattice or hexagonal lattice is realized~\cite{beth90}. Phase transition in C$_{60}$ crystal occurs at the temperature of $257 K$ and this is the first order transition. At high temperatures molecules can freely rotate whereas at low temperatures rotation is stopped and anisotropy of neighbor molecule C$_{60}$ interaction becomes important. This leads to small sharp change of distance between the nearest molecules. According to results of X-ray structure analysis~\cite{regu91} lattice constant changes from $1.4154\pm 0.0003 nm$ to $1.4111\pm 0.0003 nm$ (that is by $0.43\pm 0.06$ percent).

At low temperature, when C$_{60}$ -molecules are oriented in space, crystal lattice symmetry does not coincide with the symmetry of single molecule C$_{60}$ (icosahedral symmetry $Y$). In a unit cell of fullerite crystal lattice there are four C$_{60}$-molecules. These molecules form tetrahedron in which orientations of all molecules are the same. Tetrahedra, in their turn, form simple cubic lattice.

Fullerites are semiconductors with energy gap of $1.5-1.95 eV$~\cite{sait91}. Electrical resistivity of polycrystals C$_{60}$~\cite{regu91} monotonically changes with changing temperature and energy gap has monotonic dependence on the pressure value: an increase of energy gap under the pressure, higher than $2\times 10^{5}$ atm indicates the absence of metal-insulator transition at $p\simeq 10^{6}$atm. In the temperature region $150-400 K$ the relaxation time is temperature-independent what indicates that the carriers are localized and hopping mechanism of recombination, which includes tunneling of electrons between localized states, is realized.

It has been shown in 1991~\cite{flem91} that doping of solid fullerenes C$_{60}$ by small quantity of alkaline metal leads to formation of material with metallic type of conductivity and this material becomes superconducting at low temperatures ($T_c$ from $2.5 K$ for Na$_{2}$KC$_{60}$ to $33 K$ for RbCs$_2$C$_{60}$).
At changes of temperature, concentration of  alkaline metal, parameters and structure of lattice various phases of these compounds have been realized. In particular, at various filling $n$ ($n$ may change from 0 to 6) of lowest unoccupied molecular orbital (LUMO) the metallic, insulating or superconducting phases have been realized.
Superconductivity in doped fullerenes K$_x$C$_{60}$ has been studied theoretically in paper~\cite{zai93} and strong electron correlations have been shown to play a crucial role in superconducting state stabilization. Recently, strong electron correlation were also proven~\cite{zai11} to be responsible for superconductivity of planar carbon systems of graphene type.

Let us consider the electronic structure of C$_{60}$ in detail. In single-particle approximation, neglecting electron correlations, the following spectrum has been calculated~\cite{mani06}: 50 of 60 $p_z$ electrons of a neutral molecule fill all orbitals up to $L=4$. The lowest $L=0,1,2$ orbitals correspond to icosahedral states $a_g, t_{1u}, h_g$. All states with greater $L$ values undergo the icosahedral-field  splitting. There are 10 electrons in partially filled $L=5$ state. Icosahedral splitting ($L=5\rightarrow h_u+t_{1u}+ t_{2u}$) of this 11-fold degenerate orbital leads to the electronic configuration shown below.
Microscopic calculations and experimental data show that the completely filled highest occupied molecular orbital is of  $h_u$ symmetry, and  LUMO (3-fold degenerate) has $t_{1u}$ symmetry. At such conditions HOMO-LUMO gap appears due to icosahedral perturbation in the shell with $L=5$; energy gap found experimentally is about $1 eV$ for molecules in vacuum. A $t_{2g}$ (LUMO+1)-state, originated from $L=6$ shell, is found approximately $1 eV$ above the $t_{1u}$ LUMO.

Electron-electron correlations in C$_{60}$ are described by two main parameters: intra-molecular Coulomb repulsion $U$ and Hund's coupling $J_H$. In fullerenes the competition between intra-site Coulomb interaction (Hubbard $U$) and delocalization processes, connected with translational motion of electrons (which determines the bandwidth), causes the realization of insulator or metallic state~\cite{gunn97}.
Majority of the experimental data and theoretical calculations indicate that all materials with ions C$^{-n}_{60}$ at integer $n$ are Mott-Hubbard insulators as $U$ is quite large for all doped compounds A$_x$C$_{60}$.
Fullerides A$_x$C$_{60}$ doped with alkali metals A attract much attention of researchers due to unusual metal-insulator transition in these compounds. Only A$_3$C$_{60}$ is metallic and other phases AC$_{60}$, A$_2$C$_{60}$ and A$_4$C$_{60}$ are insulator~\cite{poir93}. This experimental fact contradicts to the results of band structure calculations (see \cite{sath92} for example) which predict purely metallic behavior. It has been noted in paper~\cite{lu94}, that for explanation of metallic behavior of Mott-Hubbard system ($x=3$ corresponds to the half-filling of the conduction band) one has to take into account a degeneracy of energy band. On the base of Gutzwiller variational approach the metal-insulator transition has been proven~\cite{lu94} to exist for all integer band fillings. It is shown that the critical value of Coulomb interaction parameter depends essentially on the band filling and degeneracy (in case of half filling $\frac{U_c}{2w}\simeq 2,8$ for double degeneracy, $\frac{U_c}{2w}\simeq 3,9$ for triple degeneracy).
The present study is devoted to investigation of Mott-Hubbard localization in electronic subsystem of fullerides with strong electron correlations within the model taking into account the orbital degeneracy of energy levels, strong Coulomb interaction and correlated hopping of electrons.

\section{The Hamiltonian of doped fulleride electronic subsystem}
Within the second quantization formalism the Hamiltonian of interacting electron systems can be written~\cite{fett71} as
\begin{eqnarray}
\label{geh_Ham}
H= -\mu \sum_{i\lambda\sigma} a_{i\lambda\sigma}^+ a_{i\lambda\sigma}
+{\sum_{ij\lambda\sigma}}'t_{ij}a_{i\lambda\sigma}^+ a_{j\lambda\sigma}+
\frac{1}{2}{\sum_{ijkl}}{\sum_{\alpha\beta\gamma\delta}}{\sum_{\sigma\sigma'}}J^{\alpha\beta\gamma\delta}_{ijkl}
a_{i\alpha\sigma}^+a_{j\beta\sigma'}^+ a_{l\delta\sigma'}a_{k\gamma\sigma},
\end{eqnarray}
where the first sum with matrix element
\begin{eqnarray}
t_{ij}=\int{d^3} r{\phi}_{\lambda i}^{*}({\bf r}-{\bf R}_{i})\times
\left[ -\frac{\hbar^2}{2m}\Delta +V^{ion}({\bf r})\right]
\phi_{\lambda i} ({\bf r}-{\bf R}_{j})
\end{eqnarray}
describes translational motion (hopping) of electrons in the crystal field $V^{ion}(\bf{r})$ and the second sum is the general expression for pair electron interactions described by matrix elements
\begin{eqnarray}
J^{\alpha\beta\gamma\delta}_{ijkl}=\int{\int{{\phi}_\alpha^{*}({\bf r}-{\bf R}_{i}){\phi}_\beta({\bf r}-{\bf R}_{j})}}\times
 \frac{e^2}{|r-r'|}{\phi}_\delta^{*}({\bf r}-{\bf R}_{l}){\phi}_\gamma({\bf r}-{\bf R}_{k})dr dr'.
 \end{eqnarray}

In the above formulae $a_{i\lambda\sigma}^{+}$, $a_{i\lambda\sigma}$ are operators of spin-$\sigma$ electron creation and annihilation in orbital state $\lambda$ on lattice site $i$, respectively, indices $\alpha$, $\beta$, $\gamma$, $\delta$, $\lambda$ denote orbital states, ${\phi}_{\lambda i}$ is wave-function in Wannier (site) representation other notation are standard.
Hamiltonian~(\ref{geh_Ham}) is essentially non-diagonal and hard to treat mathematically. The problem can be greatly simplified by neglecting the matrix elements of interaction of the third and further orders of magnitude and restrict oneself to consideration of a single orbital per site. In this way, Hamiltonian of Hubbard model and many other backbone models of strongly correlated electrons theory were derived. However, it has been shown that these models lack the possibility of description of electron-hole asymmetry, observed in real correlated electron systems. To maintain such possibility we are to consider the energy levels structure and estimate interaction parameters prior to make simplifications. Following papers~\cite{d_act00,dsdh_prb} we derive the Hamiltonial which takes into account the correlated hopping of electrons (the site-occupation dependence of hopping parameters results from taking into account the interactions with second order of magnitude matrix elements) and variety of intra-cite interactions caused by triple orbital degeneracy of LUMO in doped fullerites.
Interaction integral of  zeroth-order  magnitude is on-site Coulomb correlation (characterized by Hubbard parameter $U$):
\begin{eqnarray}
 U=\int{\int{|{\phi}_\lambda^{*}({\bf r}-{\bf R}_{i})|^2 \frac{e^2}{|r-r'|}|\phi_\lambda ({\bf r'}-{\bf R}_{i})|^2drdr'}},
\end{eqnarray}
In orbitally degenerate system, the on-site (Hund's rule) exchange integral
\begin{eqnarray}
 J_H=\int{\int{\phi}_\lambda^{*}}({\bf r}-{\bf R}_{i})\phi_{\lambda^{'}}({\bf r}-{\bf R}_{i}){e^{2}\over |{r}-{r}^{'}|}\times
\phi^{*}{_{\lambda^{'}}}({\bf r}^{'}-{\bf R}_{i})\phi_\lambda({\bf r}^{'}-{\bf R}_{i})d{\bf r}d{\bf r}^{'},
\end{eqnarray}
 is of principal importance, too. Parameter $U$ value for fullerenes have been estimated within different methods. Use of local density approximation (LDA) gives ~ 3.0 eV~\cite{ped92,antr92}.
Experimental estimation of electron repulsion energy~\cite{hett91} gives $U\simeq 2.7$ eV.

It's worth to note, that in solid state molecules are placed close enough to provide substantial screening of interaction. Calculation with screening effect took into account give $U$ 2.7 åÂ~\cite{ped92,antr92}. Combining  Auger spectroscopy and photoemission spectroscopy lead to value 1.4-1.6 eV~\cite{lof92,bruh93} for $U$.
We also note that energy cost of electron configurations with spins aligned in parallel  is considerably less than for anti-parallel alignment. Orbitally degenerate levels are filled according to Hund's rule. Experimental methods~\cite{lof92} for singlet-triplet splitting give 0.2 eV $\pm$ 0.1 eV; and in work~\cite{mart93} has the values close to 0.05 eV.
The relevant inter-site parameters are electron hopping integral and inter-site exchange coupling $J(i\lambda j{\lambda}'j\lambda i{\lambda}^{'})$.

  The resulting Hamiltonian of doped fulleride electronic subsystem reads as
 \begin{eqnarray}
&H&= -\mu \sum_{i\lambda\sigma} a_{i\lambda\sigma}^+ a_{i\lambda\sigma}+
U\sum_{i\lambda}n_{i\lambda\uparrow}n_{i\lambda\downarrow}+
\frac{U'}{2}\sum_{i\lambda\sigma}n_{i\lambda\sigma}n_{i\lambda'\bar\sigma}+
\frac{U'-J_H}{2}\sum_{i\lambda\lambda'\sigma}n_{i\lambda\sigma}n_{i\lambda'\sigma}+
\nonumber
\\
&+&{\sum_{ij\lambda\sigma}}'t_{ij}(n)a_{i\lambda\sigma}^+ a_{j\lambda\sigma}+
{\sum_{ij\lambda\sigma}}'t^{'}_{ij}\left(a_{i\lambda\sigma}^+ a_{j\lambda\sigma}n_{i\bar\lambda}+h.c. \right)+
{\sum_{ij\lambda\sigma}}'t^{''}_{ij}\left(a_{i\lambda\sigma}^+ a_{j\lambda\sigma}n_{i\lambda\bar\sigma}+h.c. \right),
\end{eqnarray}
where $n_{i\lambda\sigma}=a_{i\lambda\sigma}^{+}a_{i\lambda\sigma}$, $U'=U-2J_H$ and hopping integrals $t_{ij}(n)$, $t^{'}_{ij}, t^{''}_{ij}$ taking into account three types of correlated hopping of electrons~\cite{did97} are introduced.

\begin{figure*}% figure* for wide figure, [h] [!] to change the placement
\includegraphics[width=15cm]{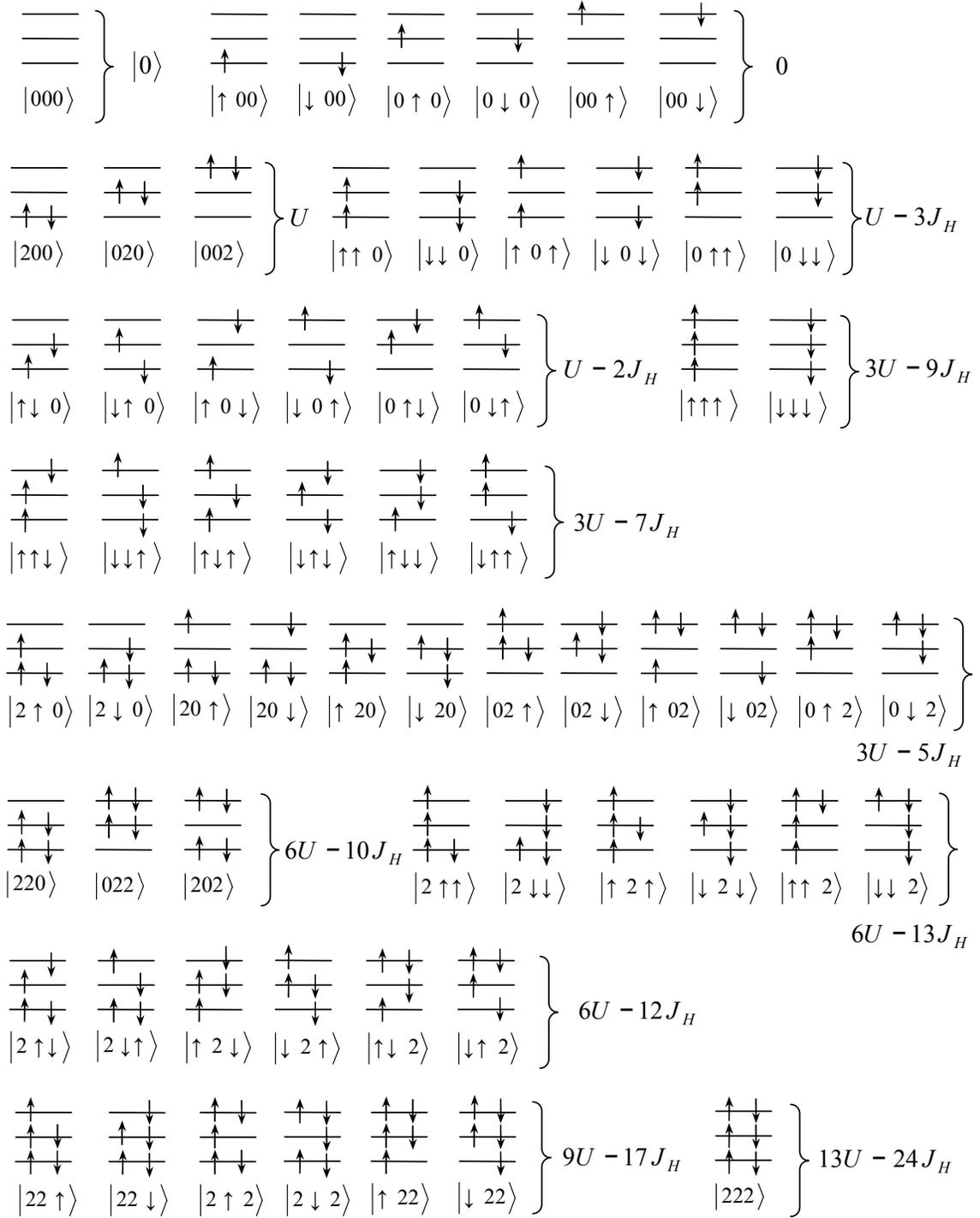}
\caption{Possible site configurations in threefold degenerate model. The first symbol in the state notation correspond to $\alpha$ orbital, the second and the third - to $\beta$ and $\gamma$ orbitals, correspondingly.}
\end{figure*}

In a model of triply degenerate band, every site can be in one of 64 configurations (see fig. 1). To pass from electron operator to Hubbard operators $X^{pl}$ of site transition from state $| l \rangle$ to state $| p \rangle$ we use relations of type
\begin{eqnarray}
\label{op}
\hat{a}_{\alpha \uparrow}^{+}&=& X^{\uparrow00,000}+X^{200,\downarrow00}+X^{\uparrow\uparrow0,0\uparrow0}+X^{\uparrow\downarrow0,0\downarrow0}
+X^{\uparrow0\uparrow,00\uparrow}+X^{\uparrow0\downarrow,00\downarrow}+X^{\uparrow20,020}+X^{\uparrow02,002}
\nonumber
\\
\nonumber
&+&
X^{2\downarrow0,\downarrow\downarrow0}+X^{20\downarrow,\downarrow0\downarrow}+X^{\uparrow\uparrow\uparrow,0\uparrow\uparrow}+X^{\uparrow\downarrow\downarrow,0\downarrow\downarrow}
+X^{2\uparrow0,\downarrow\uparrow0}+X^{20\uparrow,\downarrow0\uparrow}+X^{\uparrow\uparrow\downarrow,0\uparrow\downarrow}
+X^{\uparrow\downarrow\uparrow,0\downarrow\uparrow}
\\
\nonumber
&+&
X^{2\downarrow\downarrow,\downarrow\downarrow\downarrow}+X^{2\downarrow\uparrow,\downarrow\downarrow\uparrow}
+X^{2\uparrow\downarrow,\downarrow\uparrow\downarrow}+X^{2\uparrow\uparrow,\downarrow\uparrow\uparrow}
+X^{220,\downarrow20}+X^{\uparrow2\uparrow,02\uparrow}+X^{\uparrow2\downarrow,02\downarrow}+X^{202,\downarrow02}
\\
\nonumber
&+&X^{\uparrow\uparrow2,0\uparrow2}+
X^{\uparrow\downarrow2,0\downarrow2}+X^{\uparrow22,022}+X^{22\downarrow,\downarrow2\downarrow}
+X^{2\downarrow2,\downarrow\downarrow2}+X^{22\uparrow,\downarrow 2\uparrow}+
X^{2\uparrow2,\downarrow\uparrow2}+X^{222,\downarrow22},
\\
\nonumber
\hat{a}_{\alpha \downarrow}^{+}&=& X^{\downarrow00,000}-X^{200,\uparrow00}+X^{\downarrow\uparrow0,0\uparrow0}+X^{\downarrow\downarrow0,0\downarrow0}
+X^{\downarrow0\uparrow,00\uparrow}+X^{\downarrow0\downarrow,00\downarrow}+X^{\downarrow20,020}+X^{\downarrow02,002}
\\
\nonumber
&-&
X^{2\uparrow0,\uparrow\uparrow0}-X^{20\uparrow,\uparrow0\uparrow}+X^{\downarrow\uparrow\uparrow,0\uparrow\uparrow}+X^{\downarrow\downarrow\downarrow,0\downarrow\downarrow}
-X^{2\downarrow0,\uparrow\downarrow0}-X^{20\downarrow,\uparrow0\downarrow}+X^{\downarrow\uparrow\downarrow,0\uparrow\downarrow}
+X^{\downarrow\downarrow\uparrow,0\downarrow\uparrow}
\\
\nonumber
&-&
X^{2\uparrow\uparrow,\uparrow\uparrow\uparrow}-X^{2\uparrow\downarrow,\uparrow\uparrow\downarrow}
-X^{2\downarrow\uparrow,\uparrow\downarrow\uparrow}-X^{2\downarrow\downarrow,\uparrow\downarrow\downarrow}
-
X^{220,\uparrow20}+X^{\downarrow2\uparrow,02\uparrow}+X^{\downarrow2\downarrow,02\downarrow}-X^{202,\uparrow02}
\\
&+&X^{\downarrow\uparrow2,0\uparrow2}+
X^{\downarrow\downarrow2,0\downarrow2}+X^{\downarrow22,022}-X^{22\uparrow,\uparrow2\uparrow}
-X^{2\uparrow2,\uparrow\uparrow2}-X^{22\downarrow,\uparrow2\downarrow}-
X^{2\downarrow2,\uparrow\downarrow2}-X^{222,\uparrow22},
\end{eqnarray}
which ensure the fulfilment of anticommutation relations $\{X_{i}^{pl};X_{j}^{kt}\}=\delta_{ij}(\delta_{lk}X_{i}^{pt}+\delta_{pt}X_{i}^{kl})$, and normalizing condition $ \sum \limits_{i}X_i^p=1$, for number operators $X_i^p=X_{i}^{pl}X_{i}^{lp}$ of $|p>$-state on site $i$.
Such type of electronic operators representation is typical for models of strongly-correlated electron systems as superconducting cuprates~\cite{ovch94}, manganites~\cite{gavr11}, cobaltites~\cite{ovch11}, optical lattices~\cite{stas09,stas10}. Using the root vector notations introduced in paper~\cite{zai76} allows to obtain much more compact form of Hamiltonian in configurational representation. However, in our case number of subbands is relatively small and we use bulky but simple notations which make the projection procedure used below more transparent.

In the configurational representation the model Hamiltonian takes the form $H=H_0+T$. Here $H_0$ sums the "atomic limit" terms and the translational part may decomposed as $T=\sum \limits_{n,m}T_{nm}$, where $n,m$ serve for numbering "atomic" states. Terms $T_{nn}$ of the Hamiltonian form the energy subbands and terms of $T_{nm}$ describe the hybridization of these subbands. Different hopping integrals correspond to transitions in (or between) the different subbands. The subbands of higher-energy processes appear to be narrower due to the correlated hopping of electrons. The relative positions and overlapping of the subbands depends on the relations between the energy parameters. At integer values of electron concentration ($n=1,2,3,4,5$) in the system the metal-insulator transition is possible.

In the partial case of band filling $n=1$, strong Coulomb correlation and strong Hund's coupling (parameter $U-3J_H$ is much greater than the bandwidth, see estimations in papers~\cite{ped92,antr92}) the states with three and more electrons on the same site are excluded. Then the influence of correlated hopping can be described by three different hopping integrals. The  bare band hopping integral $t_{ij}$ is renormalized to take into account the band narrowing caused by concentration dependent correlated hopping as $t_{ij}(n)=t_{ij}(1-\tau_1n)$. This hopping integral characterizes lower Hubbard subband. Parameter $\tau_1$ is usually neglected, but it is of principle important for a consistent description of correlation effects in narrow band systems (see~\cite{d_act00, dsdh_prb} for a detailed discussion). The hopping integral for upper Hubbard subband is $\tilde{t}_{ij}(n)=t_{ij}(n)+2t'_{ij}$  and $\bar{t}_{ij}(n)=t_{ij}(n)+t'_{ij}$ describes a hybridization of lower and upper Hubbard subbands. In the following only the case $n=1$ is considered so we omit the explicit notation of concentration dependence. Then the Hamiltonian in $X-$operator representation~\cite{hubb65} has the form
\begin{eqnarray}
\label{eff_ham_fm}
&H&=H_0+\sum_{\lambda=\alpha,\beta,\gamma}\left(H_b^{(\lambda)}+H_h^{(\lambda)}\right),
\\
\nonumber
&H_0&=-\mu \sum_{i\sigma} (X_i^{\sigma00}+X_i^{0\sigma0}+X_i^{00\sigma}+
2\left( X_i^{\sigma\sigma0}+X_i^{\sigma0\sigma}+X_i^{0\sigma\sigma}\right))+
(U-3J_H)\sum_{i\sigma}\left( X_i^{\sigma\sigma0}+X_i^{\sigma0\sigma}+X_i^{0\sigma\sigma}\right),
\\
\nonumber
&H_{b}^{(\alpha)}&={\sum \limits_{ij\sigma}} ( t_{ij}X_i^{\sigma00,000}X_j^{000,\sigma00}
+
\tilde {t}_{ij}X_i^{\sigma\sigma0,0\sigma0}X_j^{0\sigma0,\sigma\sigma0}
+\tilde {t}_{ij}X_i^{\sigma0\sigma,00\sigma}X_j^{00\sigma,\sigma0\sigma}+
\\
\nonumber
&+&\tilde {t}_{ij}X_i^{\sigma\sigma0,0\sigma0}X_j^{00\sigma,\sigma0\sigma}+\tilde {t}_{ij}X_i^{\sigma0\sigma,00\sigma}X_j^{0\sigma0,\sigma\sigma0}),
\\
\nonumber
&H_{h}^{(\alpha)}&={\sum \limits_{ij\sigma}}  \bar{t}_{ij}(X_i^{\sigma00,000}X_j^{0\sigma0,\sigma\sigma0}
+X_i^{\sigma\sigma0,0\sigma0}X_j^{000,\sigma00}+
\\
\nonumber
&+&
X_i^{\sigma00,000}X_j^{00\sigma,\sigma0\sigma}+X_i^{\sigma0\sigma,00\sigma}X_j^{000,\sigma00}),
\end{eqnarray}
\begin{eqnarray*}
&H_{b}^{(\beta)}&={\sum \limits_{ij\sigma}} ( t_{ij}X_i^{0\sigma0,000}X_j^{000,0\sigma0}+
\tilde {t}_{ij}X_i^{\sigma\sigma0,\sigma00}X_j^{\sigma00,\sigma\sigma0}
+\tilde {t}_{ij}X_i^{0\sigma\sigma,00\sigma}X_j^{00\sigma,0\sigma\sigma}-
\\
&-&\tilde {t}_{ij}X_i^{\sigma\sigma0,\sigma00}X_j^{00\sigma,0\sigma\sigma}-\tilde {t}_{ij}X_i^{0\sigma\sigma,00\sigma}X_j^{\sigma00,\sigma\sigma0}),
\\
&H_{h}^{(\beta)}&={\sum \limits_{ij\sigma}} \bar{t}_{ij}(X_i^{0\sigma0,000}X_j^{00\sigma,0\sigma\sigma}
+X_i^{0\sigma\sigma,00\sigma}X_j^{000,0\sigma0}-
X_i^{0\sigma0,000}X_j^{\sigma00,\sigma\sigma0}-X_i^{\sigma\sigma0,\sigma00}X_j^{000,0\sigma0}),
\\
&H_{b}^{(\gamma)}&={\sum \limits_{ij\sigma}} ( t_{ij}X_i^{00\sigma,000}X_j^{000,00\sigma}+
\tilde {t}_{ij}X_i^{\sigma0\sigma,\sigma00}X_j^{\sigma00,\sigma0\sigma}
-\tilde {t}_{ij}X_i^{0\sigma\sigma,0\sigma0}X_j^{0\sigma0,0\sigma\sigma}+
\\
&+&\tilde {t}_{ij}X_i^{\sigma0\sigma,\sigma00}X_j^{0\sigma0,0\sigma\sigma}+\tilde {t}_{ij}X_i^{0\sigma\sigma,0\sigma0}X_j^{\sigma00,\sigma0\sigma}),
\\
&H_{h}^{(\gamma)}&=-{\sum \limits_{ij\sigma}} \bar{t}_{ij}(X_i^{00\sigma,000}X_j^{\sigma00,\sigma0\sigma}
+X_i^{\sigma0\sigma,\sigma00}X_j^{000,00\sigma}+
X_i^{00\sigma,000}X_j^{0\sigma0,0\sigma\sigma}+X_i^{0\sigma\sigma,0\sigma0}X_j^{000,00\sigma}).
\end{eqnarray*}

Green functions technique allows us to calculate the energy spectrum of the model which corresponds to the electronic subsystem  of A$_x$C$_{60}$ in the case of electron concentration $n=1$. One can rewrite the single-particle Green function $\langle\langle a_{i\lambda \sigma} | a_{j\lambda \sigma}^{+}\rangle\rangle$ on the basis of relation between electronic operators and Hubbard's X-operators:
\begin{eqnarray}
a_{p\alpha\uparrow}&=&X_p^{000,\uparrow00}+X_p^{0\uparrow0,\uparrow\uparrow0}+X_p^{00\uparrow,\uparrow0\uparrow}\equiv
 X_p^{000,\uparrow00}+Y_p,
\end{eqnarray}
where the operator $Y_p$ describes the transition processes between doubly occupied Hund's state and single occupied state. The processes involving other type of doubly occupied states, empty states, states with three or more electrons is improbable due to energy scaling.

In this way we obtain the following expression for the single electron Green function
\begin{eqnarray}
\label{green_f0}
\langle\langle a_{p\alpha\uparrow}|a_{p'\alpha\uparrow}^+\rangle\rangle=\langle\langle X_{p}^{000,\uparrow00}|X_{p'}^{\uparrow00,000}\rangle\rangle+
+ \langle\langle X_{p}^{000,\uparrow00}|Y_{p'}^{+} \rangle\rangle +\langle\langle Y_{p}|X_{p'}^{000,\uparrow00} \rangle\rangle + \langle\langle Y_{p}|Y_{p'}^{+}\rangle\rangle.
\end{eqnarray}

Equation of motion for Green function $\langle\langle X_{p}^{000,\uparrow00}|X_{p'}^{\uparrow00,000} \rangle\rangle$ has the form
\begin{eqnarray}
\label{green_f1}
\nonumber
(E+\mu)\langle\langle X_{p}^{000,\uparrow00}|X_{p'}^{\uparrow00,000} \rangle\rangle
&&=\delta_{pp'}\frac{X_p^{000}+X_p^{\uparrow00}}{2\pi}
+\langle\langle [X_{p}^{000,\uparrow00};\sum_\lambda{H_b^{(\lambda)}}]|X_{p'}^{\uparrow00,000} \rangle\rangle
\\
&&+ \langle\langle[X_{p}^{000,\uparrow00};\sum_\lambda{H_h^{(\lambda)}}]|X_{p'}^{\uparrow00,000} \rangle\rangle
\end{eqnarray}
and equation of motion for Green function $\langle\langle Y_{p}|X_{p'}^{000,\uparrow00} \rangle\rangle$ -
\begin{eqnarray}
\label{green_f2}
\nonumber
(E+\mu-U+3J_H)\langle\langle Y_{p}|X_{p'}^{000,\uparrow00} \rangle\rangle=
\langle\langle [Y_{p};\sum_\lambda{H_b^{(\lambda)}}]|X_{p'}^{\uparrow00,000} \rangle\rangle
+ \langle\langle Y_{p};\sum_\lambda{H_h^{(\lambda)}}]|X_{p'}^{\uparrow00,000} \rangle\rangle.
\end{eqnarray}

To obtain closed system of equations for Green functions $\langle\langle X_{p}^{000,\uparrow00}|X_{p'}^{\uparrow00,000} \rangle\rangle$  and $\langle\langle Y_{p}|X_{p'}^{\uparrow00,000} \rangle\rangle$ we use the projection procedure similar to the work~\cite{did97}:
\begin{eqnarray}
\label{project}
[X_{p}^{000,\uparrow00};\sum_\lambda{H_b^{(\lambda)}}]&=&\sum_{i}\varepsilon_{pi}^bX_{i}^{000,\uparrow00};
\\
\nonumber
[X_{p}^{000,\uparrow00};\sum_\lambda{H_h^{(\lambda)}}]&=&\sum_{i}\varepsilon_{pi}^hY_i;
\\
\nonumber
[Y_{p};\sum_\lambda{H_b^{(\lambda)}}]&=&\sum_{i}\tilde \varepsilon_{pi}^bY_i;
\\
\nonumber
[Y_{p};\sum_\lambda{H_h^{(\lambda)}}]&=&\sum_{i}\tilde \varepsilon_{pi}^hX_{i}^{000,\uparrow00}.
\end{eqnarray}
As a result after Fourier transformation we obtain the Green function in the form:
\begin{eqnarray}
\label{green_f}
\langle\langle X_{i}^{000,\uparrow00}|X_{j}^{\uparrow00,000} \rangle\rangle_{\bf k}=\frac{X^{000}+X^{\uparrow00}}{2\pi}\times
\frac{E+\mu-U+3J_H-\tilde\varepsilon^b({\bf k})}{(E-E_1({\bf k}))(E-E_2({\bf k}))},
\end{eqnarray}
where the quasi-particle energy spectrum
\begin{eqnarray}
\label{spectrum}
E_{1,2}({\bf k})=-\mu+\frac{U-3J_H}{2}+\frac{\varepsilon^b({\bf k})+\tilde\varepsilon^b({\bf k})}{2}\mp
\frac{1}{2}\sqrt{(U-3J_H-\varepsilon^b({\bf k})+\tilde\varepsilon^b({\bf k}))^2+4\varepsilon^h({\bf k})\tilde\varepsilon^h({\bf k})}.
\end{eqnarray}
In the absence of orbital order the energy spectrum for $\beta$ and $\gamma$ electrons is the same as for $\alpha$ electrons.

The non-operator coefficients $\varepsilon^b({\bf k}),\tilde\varepsilon^b({\bf k}),\varepsilon^h({\bf k}),\tilde\varepsilon^h({\bf k})$ one can obtain by the anticommutation of Eq.(\ref{project}) with basis operators $X_{i}^{000,\uparrow00}$ and $Y_i^+$ and following replacement of operators by $c$-numbers (see in this connection~\cite{d_act00}).
\begin{eqnarray*}
\label{eps}
&&\varepsilon_{\bf k}^b=\frac{1}{C_1}[t_{\bf k}(
\langle X_p^{000}(X_{p'}^{000}+X_{p'}^{\uparrow00})\rangle+
+\langle X_p^{\uparrow 00}(X_{p'}^{000}+X_{p'}^{\uparrow00})\rangle+
\langle X_p^{\downarrow 00,\uparrow 00} X_{p'}^{\uparrow 00,\downarrow 00}\rangle+
\langle X_p^{0\uparrow 0,\uparrow 00} X_{p'}^{\uparrow 00,0\uparrow 0}\rangle+
\\
\nonumber
&&+
\langle X_p^{0\downarrow 0,\uparrow 00} X_{p'}^{\uparrow 00,0\downarrow 0}\rangle+
+\langle X_p^{00\uparrow,\uparrow 00} X_{p'}^{\uparrow 00,00\uparrow}\rangle+
\langle X_p^{00\downarrow,\uparrow 00} X_{p'}^{\uparrow 00,00\downarrow}\rangle
)-
-\tilde{t}_{\bf k}(\langle X_p^{\uparrow \uparrow0,000} X_{p'}^{000,\uparrow \uparrow0}\rangle+
\langle X_p^{\uparrow0\uparrow,000} X_{p'}^{000,\uparrow0\uparrow}\rangle)],
\end{eqnarray*}
\begin{eqnarray*}
&&\varepsilon_{\bf k}^h=\frac{1}{C_2}
\bar{t}_{\bf k}[\langle (X_p^{000}+X_{p}^{\uparrow00})\times
(X_{p'}^{0\uparrow0}+X_{p'}^{00\uparrow}+
X_{p'}^{\uparrow\uparrow0}+X_{p'}^{\uparrow0\uparrow}
)\rangle+
\langle X_p^{0\uparrow0,\uparrow 00} X_{p'}^{\uparrow 0\uparrow,0\uparrow\uparrow}\rangle+
\langle X_{p'}^{\uparrow\uparrow 0,000} X_{p}^{000,\uparrow\uparrow 0}\rangle-
\\
\nonumber
&&-\langle X_{p'}^{0\uparrow0,\uparrow 00} X_{p}^{\uparrow 00,0\uparrow0}\rangle
-\langle X_p^{00\uparrow,\uparrow 00} X_{p'}^{\uparrow 00,00\uparrow}\rangle+
\langle X_{p'}^{\uparrow0\uparrow,000} X_{p}^{000,\uparrow0\uparrow}\rangle-
\langle X_{p}^{00\uparrow,\uparrow 00} X_{p'}^{\uparrow\uparrow0,0\uparrow\uparrow}\rangle],
\end{eqnarray*}
\begin{eqnarray*}
&&\tilde{\varepsilon}_{\bf k}^b=-\frac{t_{\bf k}}{C_2}
[\langle X_{p'}^{\uparrow\uparrow0,000} X_{p}^{000,\uparrow\uparrow0}\rangle
+\langle X_{p'}^{\uparrow0\uparrow,000} X_{p}^{000,\uparrow0\uparrow}\rangle
]+
\\
&&+\frac{\tilde{t}_{\bf k}}{C_2}
[\langle X_p^{0\uparrow0}+X_p^{\uparrow\uparrow0}+X_p^{00\uparrow}+X_p^{\uparrow0\uparrow}\rangle
\times\langle X_{p'}^{0\uparrow0}+X_{p'}^{\uparrow\uparrow0}+X_{p'}^{00\uparrow}+X_{p'}^{\uparrow0\uparrow}\rangle+
\langle X_p^{0\uparrow0,\uparrow 00} X_{p'}^{\uparrow 00,0\uparrow0}\rangle+
\langle X_{p}^{0\uparrow\uparrow,\uparrow0\uparrow} X_{p}^{\uparrow0\uparrow,0\uparrow\uparrow}\rangle-
\\
\nonumber
&&-\langle X_{p}^{0\uparrow0,\uparrow 00} X_{p'}^{\uparrow 0\uparrow,0\uparrow\uparrow}\rangle-
\langle X_p^{0\uparrow\uparrow,\uparrow 0\uparrow} X_{p'}^{\uparrow 00,0\uparrow0}\rangle+
\langle X_{p}^{00\uparrow,\uparrow00} X_{p'}^{\uparrow00,00\uparrow}\rangle+
\langle X_{p}^{0\uparrow\uparrow,\uparrow \uparrow0} X_{p'}^{\uparrow \uparrow0,0\uparrow\uparrow}\rangle+
\\
\nonumber
&&+\langle X_p^{00\uparrow,\uparrow 00} X_{p'}^{\uparrow\uparrow0,0\uparrow\uparrow}\rangle+
\langle X_{p}^{0\uparrow\uparrow,\uparrow\uparrow0} X_{p'}^{\uparrow00,00\uparrow}\rangle],
\end{eqnarray*}
\begin{eqnarray*}
\label{tildeeps_h}
\nonumber
&&\tilde{\varepsilon}_{\bf k}^h=-\frac{\bar{t}_{\bf k}}{C_1}
[\langle (X_{p'}^{000}+X_{p'}^{\uparrow00})\times
(X_{p}^{\uparrow\uparrow0}+X_{p}^{\uparrow0\uparrow}+
X_{p}^{0\uparrow0}+X_{p}^{00\uparrow}
)\rangle+
\langle X_p^{0\uparrow\uparrow,\uparrow 0\uparrow} X_{p'}^{\uparrow 00,0\uparrow0}\rangle-
\langle X_{p}^{0\uparrow0,\uparrow00} X_{p'}^{\uparrow00,0\uparrow 0}\rangle+
\\
\nonumber
&&+\langle X_{p'}^{\uparrow\uparrow0,000} X_{p}^{000,\uparrow\uparrow0}\rangle+
+\langle X_p^{00\uparrow,\uparrow 00} X_{p'}^{\uparrow 00,00\uparrow}\rangle-
\langle X_{p'}^{\uparrow0\uparrow,000} X_{p}^{000,\uparrow0\uparrow}\rangle
\langle X_{p}^{0\uparrow\uparrow,\uparrow \uparrow0} X_{p'}^{\uparrow00,00\uparrow}\rangle],
\end{eqnarray*}
where $C_1=\langle X_p^{000}+X_p^{\uparrow00}\rangle$, $C_2=\langle X_p^{0\uparrow0}+X_p^{00\uparrow}+X_p^{\uparrow\uparrow0}+X_p^{\uparrow0\uparrow}\rangle$. It is worth to note that in the partial case of band filling $n=1$ and strong Coulomb correlation we work with reduced Hilbert space of electronic states, so $C_1+C_2=1$.

Let us denote the concentration of empty lattice sites by $e$, concentration of singly occupied sites with spin $\sigma$ electron in orbital state $\lambda$ by $s_{\lambda\sigma}$, Hund's doublons concentration by $d_\sigma$, Hubbard doublons by $d_2$ and non-Hund doublons by $\tilde{d}$. In a paramagnetic state $s_{\lambda\sigma}=s$, $d_{\sigma}=d$. For the case of strong Hund's coupling the high energy doublon configurations are excluded, $d_2=\tilde{d}=0$. We can utilize the completeness condition for the $X$-operator set to have constraint $e+6s+6d=1$, which, at condition $e=6d$, leads to the equation
\begin{eqnarray}
s=\frac{1-12d}{6}.
\end{eqnarray}

Finally in the paramagnetic case at $n=1$ we obtain
\begin{eqnarray}
\label{coef}
&&\varepsilon^b=\frac{216d^2-12d+1}{24d+1}t_{\bf k}+\frac{72d^2}{24d+1}\tilde t_{\bf k};
\\
&&\varepsilon^h=\bar{t}_{\bf k}\frac{7d-12d^2}{1-6d},
\\
&&\tilde\varepsilon^b=t_{\bf k}\frac{36d^2}{1-6d}+\frac{\tilde{t}_{\bf k}}{2(1-6d)},
\\
&&\tilde\varepsilon^h=t_{\bf k}\frac{24d+1-216d^2}{3(24d+1)},
\end{eqnarray}

In this way, the energy spectrum depends on the concentration of doublons $d$ (through the dependence of non-operator coefficients). The doublon concentration is determined by the condition
\begin{eqnarray}
6d={1\over 2N}\sum_{\bf k}{\left(\frac{A_e(\bf k)}{\exp(\frac{E_1(\bf k)}{kT}+1)}+\frac{B_e(\bf k)}{\exp(\frac{E_2(\bf k)}{kT}+1)}\right)},
\end{eqnarray}
where
\begin{eqnarray}
\nonumber
&&A_e({\bf k})={1\over 2}\left(1+\frac{U-3J_H+\tilde{\varepsilon}^b-\varepsilon^b}
{\sqrt{(U-3J_H-\varepsilon^b+\tilde{\varepsilon}^b)^2+4\varepsilon^h\tilde{\varepsilon}^h}}\right),
\\
&&B_e({\bf k})=1-A_e({\bf k}).
\end{eqnarray}
Using the model rectangular density of states at zero temperature one obtains
\begin{eqnarray}
6d={1\over 4w}\int^{w}_{-w}{\frac{A_e(\varepsilon)\Theta(-E_1(\varepsilon))}{E-E_1(\varepsilon)}d\varepsilon}
{{1\over 4w}\int^{w}_{-w}\frac{B_e(\varepsilon)\Theta(-E_2(\varepsilon))}{E-E_2(\varepsilon)}d\varepsilon},
\end{eqnarray}
here $\Theta(-E(\varepsilon))$ is Heaviside theta-function.
Solving this equation numerically we obtain the doublon concentration as function of the model parameters.
To study a metal-insulator transition (MIT)~\cite{mott90,edv95,geb97} we apply the gap criterion
\begin{eqnarray}
\Delta E=E_2(-w)-E_1(w)=0.
\end{eqnarray}
In the point of MIT the polar states (holes and doublons) concentrations equals zero. Thus, for the non-operators coefficients we have $\varepsilon^b=t_{\bf k}$, $\varepsilon^h=0$, $\tilde{\varepsilon}^b=\frac{\tilde{t}_{\bf k}}{2}$, $\tilde{\varepsilon}^h=\frac{\tilde{t}_{\bf k}}{3}$, and for the energy gap we have the equation
\begin{eqnarray}
\label{mit_crit}
\Delta E=U-3J_H-\tilde{w}-w.
\end{eqnarray}
Here $w=z|t|(1-\tau_1)$ and $\tilde{w}=z|t|(1-\tau_1)(1-2\tau)$ are the halfbandwidths of the lower and upper subbands, respectively, $z$ is the number of nearest neighbours to a site, $|t|$ is the magnitude of bare nearest-neighbour hopping integral,  $\tau_1, \tau=\frac{t'_{ij}}{|t_{ij}|}$ are the correlated hopping parameters. From the equation~(\ref{mit_crit}) one obtains that the critical value of the intra-cite Coulomb interaction parameter equals the sum of quasiparticle subbands halfbandwiths.

Analysis of the expression~(\ref{mit_crit}) allows explaining the differences of electrical characteristics (insulator or metallic state realisation) depending on the correlated hopping strength.

The correlated hopping influence substantially on electrical characteristics of narrow band material with three-fold orbital degeneracy of the energy levels. Both the filling of the sites involved into the hopping processes (through the correlated hopping of the first type) and the neighbor sites  (through the second type correlated hopping), can lead to appearance of the gap in energy spectrum and stabilization of the insulator state. The energy gap, however, opens at relatively large increase of correlated hopping parameters which can not be achieved in a compound by change of external conditions only. Such critical increase of parameters $\tau_1$ and $\tau$ can be realized at doping. A distinct picture is observed at the change of intra-site Coulomb interaction parameter. At increase of $(U-3J_H)/ w$ over a critical value (dependent on the correlated hopping strength) the energy gap occurs and the metal-insulator transition takes place. The critical value for the partial case of the model when the quasiparticle subbands have the same widths (in absence of the correlated hopping), is $(U-3J_H)/ w=2$ which corresponds to the result of works~\cite{did97,did98} for non-degenerated Hubbard model.

\section{Conclusions}
Within the variant of triple orbitally degenerate model of the electronic subsystem of a doped fulleride compound considered above not only the on-site Coulomb correlations but also additional interactions of principal importance, namely the correlated hopping, can be introduced and analyzed. The use of Hubbard X-operators representation appears to be useful to exclude from consideration the parts of Hilbert space which are irrelevant at particular band filling. The ground state metal-insulator transition in the triply degenerate model of partially-filled doped fulleride band takes place at moderate values of the correlation parameter which in this case is a combination of on-site Coulomb repulsion energy, Hund's rule coupling and electron hopping parameters. The correlated hopping of electrons leads to further localization of current carriers. The influence of the correlated hopping is substantial and makes the estimation of the model parameters from the available spectroscopic data ambiguous. The problem can be resolved by the additional spectroscopic experiments with use of external pressure. In this case the reasonable estimates could be obtained using the fact that in distinction from the on-site parameters, the correlated hopping parameters must be pressure-dependent. The metal-insulator transition described above can be realized~\cite{deg98, sac01} in the doped fulleride compound under the external pressure.

\end{document}